\documentclass[preprint,showpacs,preprintnumbers]{revtex4}
\usepackage{amsmath}
\usepackage{graphicx}
\usepackage{dcolumn}
\usepackage{bm}
\usepackage{hyperref}
\usepackage{amsmath}
\usepackage{amsfonts}
\usepackage{amssymb}

\input{tcilatex}
\begin{document}

\title{Quantized Gravitational Radiation from Black Holes and other Macro
Holeums in the Low Frequency Domain}
\author{A. L. Chavda$^{a}$, L. K. Chavda$^{b}$}
\email{$^{a}$\ a01.l.chavda@gmail.com, $^{b}$\ holeum@gmail.com}
\affiliation{$^{a}$ Physics Department, South Gujarat University, Surat - 395007, $^{b}$
49, Gandhi Society, City Light Road, Via Parle Point, Surat - 395007.}
\date{\today }

\begin{abstract}
A gravitational bound state, called a Macro Holeum, is created from a very
large number of microscopic black holes of primordial or a non-primordial
origin. All of them undergo orbital motion, under the action of gravity,
around their common center of mass. Four classes of Macro Holeum emerge: H,
BH, HH and LH. The latter is a massless bundle of gravitational energy
moving at the speed of light. The others are Dark Matter objects. BH emits
Hawking radiation but the others do not. The presently existing black holes
are identified with BH. All, except LH, emit gravitational radiation due to
internal quantum transitions. Simple analytic formulae are derived for the
energy eigenvalues, mass, radius, density and the frequency of the
gravitational radiation emitted by Macro Holeums in terms of just two
parameters which can be determined from the gravitational spectra. We
predict that black holes have internal structure and that they and other
Macro Holeums, having masses in the range 57 solar masses to 870 solar
masses, would emit hydrogen-like gravitational radiation in the LIGO
frequency range in the form of band or line spectra with a considerable
overlapping of the bands. This will be superposed on a uniform background
radiation coming from Macro Holeums at large distances greater than 100 Mpc.
\end{abstract}

\maketitle
\tableofcontents

\section{Introduction}

The nature of Dark Matter (DM) is of great current interest. Compelling
theoretical and observational evidence has built up over the past two
decades that the visible matter in the universe makes up a very small
fraction of the total matter composition of the universe \cite{K.A.Olive}.
It is now believed that the universe is composed of approximately 73\% dark
energy (DE), 23\% DM, and about 4\% visible matter \cite%
{M.J.Rees,D.B.Cline,W.Boer}. Consequently, the nature and identity of DM has
become one of the most challenging problems facing modern cosmology. Several
candidates for DM particles have been proposed, which include Standard Model
neutrinos \cite{G.Bertone.et.al,L.Bergstrom}, Sterile neutrinos \cite%
{S.Dodelson.et.al,G.Bertone.et.al}, Axions \cite{G.Bertone.et.al,P.Sikivie2}%
, Supersymmetric particles \cite%
{G.Bertone.et.al,T.Falk.et.al,J.L.Feng.et.al,K.Choi.et.al,L.Covi.et.al,L.Covi.et.al2}%
, dark matter from Little Higgs models \cite%
{N.Arkani-Hamed.et.al,N.Arkani-Hamed.et.al2,N.Arkani-Hamed.et.al3,N.Arkani-Hamed.et.al4,H.Cheng.et.al}%
, Kaluza-Klein dark matter \cite{H.Cheng.et.al2,A.Barrau.et.al.2},
WIMPZILLAs \cite{E.Kolb.et.al}, Cryptons \cite%
{J.R.Ellis.et.al,J.R.Ellis.et.al.2}, primordial black holes \cite%
{A.Dolgov.et.al,D.Blais.et.al,N.Ashfordi.et.al}, WIMPs \cite%
{R.Bernabei.et.al}, and super-heavy X-particles \cite%
{C.Barbot,C.Barbot.et.al,C.Barbot.et.al.2,Sanghyeon.Chang.et.al,Karim.Benakli.et.al,K.Hamaguchi.et.al,K.Hamaguchi.et.al2,K.Hamaguchi.et.al3,C.Coriano.et.al,D.J.Chung.et.al,V.Berezinsky}%
. Galaxies have been observed to have DM halos that contain far more matter
than their visible regions \cite{F.Combes,H.Hoekstra.et.al,A.Burkert}. Most
galaxies are not dominated by DM inside their optically visible disks \cite%
{F.Combes,H.Hoekstra.et.al,A.Burkert}, which suggests that DM has properties
that segregate it from visible matter.

There has been no unequivocal detection of DM particles so far. Meanwhile,
we have proposed the existence of a new form of DM which we call Holeum \cite%
{Chavda.Chavda}, based on the copious production of primordial black holes
(PBHs) that is believed to have occurred in the early universe \cite{Ya.B.
Zeldovich,S.W.Hawking4,M.Yu.Khlopov.et.al,A.G.Polnarev.et.al,S.W.Hawking5,J.Garriga.et.al,R.Caldwell.et.al,J.H.MacGibbon.et.al,M.Crawford.et.al,S.W.Hawking6,H.Kodama.et.al,D.La.et.al,I.G.Moss.et.al,H.Kodama.et.al.2,K.Maeda.et.al,R.V.Konoplich.et.al,M.Yu.Khlopov.et.al.2,V.I.Dokuchaev.et.al,S.G.Rubin.et.al,S.G.Rubin.et.al.2,M.Yu.Khlopov.et.al.3}%
. The Holeum is a quantized bound state of two PBHs. It is the gravitational
analogue of a hydrogen atom. Its atomic transitions give off extremely
high-frequency gravitational radiation which cannot be detected as yet. The
Holeum is therefore a DM particle. Since a vast quantity of microscopic
black holes is believed to have been produced in the early universe, Holeums
could constitute an important component of the DM present in the universe
today. A segregation property puts the Holeums in the Galactic Halos and the
Domain Walls, if the latter exist. Recently, we have suggested that Holeums
may be the progenitors of cosmic rays of all energies including the Ultra
High Energy Cosmic Rays (UHECR) \cite{Chavda.Chavda2}. Holeums have also
been suggested to be responsible for a class of short duration gamma ray
bursts (GRB) \cite{Al.Dallal}. Thus, Holeum theory holds a possibility of
providing a unified explanation for a number of cosmological phenomena.
However, a simple Holeum consisting of just two microscopic black holes
emits gravitational radiation in an extremely high frequency range which is
not detectable in the foreseeable future. Therefore, in this paper, we
consider the formation of a Macro Holeum consisting of a very large number
of microscopic black holes in the hope that such a Macro Holeum will radiate
gravitational energy in the low-frequency range accessible to the Laser
Interferometer Gravity-Wave Observatory (LIGO) which is now operational.
Another innovation we are introducing is motivated by the following fact:
The astrophysical black holes existing today are a part of DM. Some of them
may be of primordial origin while the others have been created by the
gravitational collapse of some stars. Any theory of DM must necessarily
include them also. To this end, we are enlarging the scope of this theory by
considering not only the PBHs for the formation of a Macro Holeum but also
the nonprimordial microscopic black holes. Hopefully, this will enable us to
include all the presently existing astrophysical black holes in the ranks of
Macro Holeums leading to interesting and testable predictions.

In section 2, we state the assumptions and build a Macro Holeum in which the
microscopic black holes orbit the center of mass of the system under the
action of gravity. We derive simple formulae for the various properties of a
Macro Holeum. We find that Macro Holeums come in four distinct classes,
based on their constituent masses. We enumerate the classes of Macro Holeums
and the ways to~determine their properties from their gravitational
frequency spectrum. We identify the presently existing astrophysical black
holes with one class of Macro Holeum and present its consequences. We
comment on the nature of the energy and the frequency spectra and on the
dependence of the density of a Macro Holeum on its parameters. We determine
the ball-park range of the masses of the Macro Holeums that will emit the
gravitational radiation in the low-frequency domain accessible to LIGO. We
also study the unique properties of a Macro Holeum in its ground state. In
section 3, we present some implications of this theory. Discussions and
conclusions are presented in section 4.

\section{Derivation of the Properties of a Macro Holeum}

\subsection{Introduction}

We would like to study the creation of a bound state of a very large number
of microscopic black holes brought about by the gravitational interaction.
We call it a Macro Holeum. The term "microscopic black holes" will be used
as a general term that includes both the primordial and the non-primordial
varieties. We will use the acronym PBH to refer specifically to the former
variety. The term "black hole" will refer to all presently existing black
holes irrespective of how they were formed.

While building Macro Holeums from microscopic black holes, it is instructive
to recall the original quark model which built hadrons such as protons,
neutrons, pi-mesons, etc. from quarks. The quark model posited the existence
of three quarks, specified their quantum numbers and specified the recipe
for making baryons as bound states of three quarks and mesons as bound
states of a quark and an antiquark. It made no hypothesis about the origin
of the quarks, nor about how and when these hadrons were actually created
from the quarks. We take a similar stance here: We make a hypothesis neither
about the origin of the microscopic back holes, nor about how and when Macro
Holeums were formed. We assume that a Macro Holeum is a bound state of $k$
microscopic black holes, all of which revolve around their common center of
mass under the action of gravity. Here $k$ is a very large but finite
positive integer. It may be large enough to give a Macro Holeum a solar mass
or even millions of solar masses.

\subsection{The exact formulae for the properties}

Let us consider a Holeum that is a bound state of two microscopic black
holes of masses $m_{1}$ and $m_{2}$ revolving~around their center of mass
under the action of gravity. From \cite{Chavda.Chavda}, the energy values of
the Holeum are given by%
\begin{equation}
E_{2}=-\frac{m_{1}m_{2}c^{2}\alpha_{2}^{2}}{2\left( m_{1}+m_{2}\right)
n_{2}^{2}}  \label{1}
\end{equation}
where $\alpha_{2}$ is the gravitational coupling constant for the
interaction between masses $m_{1}$ and $m_{2}$, given by%
\begin{equation}
\alpha_{2}=\frac{m_{1}m_{2}}{m_{p}^{2}}  \label{2}
\end{equation}
where%
\begin{equation}
m_{p}=\left( \frac{\hbar c}{G}\right) ^{\frac{1}{2}}  \label{3}
\end{equation}
is the Planck mass, $\hbar$ is the Planck's constant divided by $2\pi$, $c$
is the speed of light in vacuum, and $G$ is Newton's universal constant of
gravity. $n_{2}$ is the principal quantum number of the bound state. From 
\cite{Chavda.Chavda}, the most probable radius of the Holeum is given by%
\begin{equation}
r_{2}=\frac{\left( R_{1}+R_{2}\right) \pi^{2}n_{2}^{2}}{16\alpha_{2}^{2}}
\label{4}
\end{equation}
where%
\begin{equation}
R_{i}=\frac{2m_{i}G}{c^{2}}  \label{5}
\end{equation}
is the Schwarzschild radius of the microscopic black hole of mass $m_{i}$.
The mass of the bound state is given by%
\begin{equation}
M_{2}=m_{1}+m_{2}+\frac{E_{2}}{c^{2}}  \label{6}
\end{equation}
We shall call this bound state of two microscopic black holes a di-Holeum.
Now consider a bound state of the di-Holeum and a microscopic black hole of
mass $m_{3}$. We call it a tri-Holeum. Its energy eigen values are given by%
\begin{equation}
E_{3}=-\frac{m_{3}M_{2}c^{2}\alpha_{3}^{2}}{2\left( m_{3}+M_{2}\right)
n_{3}^{2}}  \label{7}
\end{equation}
where%
\begin{equation}
\alpha_{3}=\frac{m_{3}M_{2}}{m_{p}^{2}}  \label{8}
\end{equation}
The bound state radius of the tri-Holeum is given by%
\begin{equation}
r_{3}=\frac{\left( m_{3}+M_{2}\right) G\pi^{2}n_{3}^{2}}{8c^{2}\alpha
_{3}^{2}}  \label{9}
\end{equation}
With the help of eq. (\ref{6}) this may be written as%
\begin{equation}
r_{3}=\frac{\left( R_{1}+R_{2}+R_{3}+\frac{E_{2}}{c^{2}}\right)
\pi^{2}n_{3}^{2}}{16\alpha_{3}^{2}}  \label{10}
\end{equation}
The mass of the tri-Holeum is given by%
\begin{equation}
M_{3}=m_{3}+M_{2}+\frac{E_{3}}{c^{2}}  \label{11}
\end{equation}
Continuing in this manner we arrive at a $k-$Holeum, also called a Macro
Holeum, consisting of $k$ microscopic black holes. Its energy values, bound
state radius, and bound state mass are given, respectively, by%
\begin{equation}
E_{k}=-\frac{m_{k}^{3}c^{2}\left[ m_{1}+m_{2}+...+m_{k-1}+\frac{\left(
E_{2}+E_{3}+...+E_{k-1}\right) }{c^{2}}\right] ^{3}}{2m_{p}^{4}n_{k}^{2}%
\left[ m_{1}+m_{2}+...+m_{k}+\frac{\left( E_{2}+E_{3}+...+E_{k-1}\right) }{%
c^{2}}\right] }  \label{12}
\end{equation}%
\begin{equation}
r_{k}=\frac{\left( R_{1}+R_{2}+...+R_{k}\right) \left[ 1+\frac{%
E_{2}+E_{3}+...+E_{k-1}}{\left( m_{1}+m_{2}+...+m_{k}\right) c^{2}}\right]
\pi^{2}n_{k}^{2}}{\left( \frac{m_{k}}{m_{p}}\right) ^{2}\left[ \frac {%
m_{1}+m_{2}+...+m_{k-1}}{m_{p}}\right] ^{2}\left[ 1+\frac{%
E_{2}+E_{3}+...+E_{k-1}}{\left( m_{1}+m_{2}+...+m_{k-1}\right) c^{2}}\right]
^{2}}  \label{13}
\end{equation}%
\begin{equation}
M_{k}=\left( m_{1}+m_{2}+...+m_{k}+\frac{E_{2}+E_{3}+...+E_{k}}{c^{2}}\right)
\label{14}
\end{equation}
The gravitational coupling constant $\alpha_{k}$ for the interaction between
a Macro Holeum containing $k-1$ microscopic black holes and the microscopic
black hole of mass $m_{k}$ is given by%
\begin{equation}
\alpha_{k}=\frac{m_{k}M_{k-1}}{m_{p}^{2}}  \label{15}
\end{equation}
Using eq. (\ref{14}) this may be rewritten as%
\begin{equation}
\alpha_{k}=\frac{m_{k}\left( m_{1}+m_{2}+...+m_{k-1}+\frac{\left(
E_{2}+E_{3}+...+E_{k-1}\right) }{c^{2}}\right) }{m_{p}^{2}}  \label{16}
\end{equation}
For the sake of simplicity we now consider an equal-mass case: $%
m_{1}=m_{2}=...=m_{k}=m$, say. Then eqs. (\ref{12}) - (\ref{16}) reduce,
respectively, to%
\begin{equation}
E_{k}=-\frac{\left( k-1\right) ^{3}mc^{2}\alpha_{g}^{2}}{2kn_{k}^{2}}\left[ 
\frac{g_{k-1}\left( \alpha_{g},n_{2},n_{3},...,n_{k-1}\right) ^{3}}{%
f_{k-1}\left( \alpha_{g},n_{2},n_{3},...,n_{k-1}\right) }\right]  \label{17}
\end{equation}%
\begin{equation}
r_{k}=\frac{kR\pi^{2}n_{k}^{2}}{16\alpha_{g}^{2}\left( k-1\right) ^{2}}\left[
\frac{f_{k-1}\left( \alpha_{g},n_{2},n_{3},...,n_{k-1}\right) }{%
g_{k-1}\left( \alpha_{g},n_{2},n_{3},...,n_{k-1}\right) ^{2}}\right]
\label{18}
\end{equation}%
\begin{equation}
M_{k}=kmg_{k}\left( \alpha_{g},n_{2},n_{3},...,n_{k}\right)  \label{19}
\end{equation}%
\begin{equation}
\alpha_{k}=\left( k-1\right) \alpha_{g}g_{k-1}\left(
\alpha_{g},n_{2},n_{3},...,n_{k-1}\right)  \label{20}
\end{equation}
where%
\begin{equation}
\alpha_{g}=\left( \frac{m}{m_{p}}\right) ^{2}  \label{21}
\end{equation}
Here $n_{j}$ is the principal quantum number of the microscopic black hole
of mass $m_{j}$; $j=2,3,...k$. From eqs. (\ref{12}) and (\ref{13}) as well
as from eqs. (\ref{17}) and (\ref{18}) we see that whereas $E_{k}$ and $%
r_{k} $ depend strongly upon $n_{k}$, the principal quantum number of the
outermost microscopic black hole, their dependence on the rest of the
principal quantum numbers is quite symmetric. None of the latter is singled
out the way $n_{k}$ is. Hence for the sake of convenience we shall describe
the $k-$Holeum in terms of a "valence" microscopic black hole, namely the
outermost one and and the "core" microscopic black holes, namely, the rest
of them, $j=2,3,...k-1.\ f_{k-1}$ and $g_{k-1}$ are given by%
\begin{equation}
f_{k-1}\left( \alpha_{g},n_{2},n_{3},...,n_{k-1}\right) =1+\frac{%
E_{2}+E_{3}+...+E_{k-1}}{kmc^{2}}  \label{22}
\end{equation}%
\begin{equation}
g_{k-1}\left( \alpha_{g},n_{2},n_{3},...,n_{k-1}\right) =1+\frac{%
E_{2}+E_{3}+...+E_{k-1}}{\left( k-1\right) mc^{2}}  \label{23}
\end{equation}
Note that we must have at least $k=2$ for a bound state formation. Thus in
eqs. (\ref{22}) and (\ref{23}), we must define $f_{1}=g_{1}=1$. Note that
the Schwarzschild radius, $R_{k}$, of the $k-$Holeum containing $k$
microscopic black holes is given by%
\begin{equation}
R_{k}=\frac{2M_{k}G}{c^{2}}=kRg_{k}\left(
\alpha_{g},n_{2},n_{3},...,n_{k}\right)  \label{24}
\end{equation}
where $M_{k}$ is given by eq. (\ref{19}). From eqs. (\ref{18}) and (\ref{24}%
) we get the ratio of the Schwarzschild radius $R_{k}$ of the $k-$Holeum to
its bound state radius $r_{k}$ as follows:%
\begin{equation}
\frac{R_{k}}{r_{k}}=\frac{16\alpha_{g}^{2}\left( k-1\right) ^{2}}{\pi
^{2}n_{k}^{2}}\frac{g_{k}\left( \alpha_{g},n_{2},n_{3},...,n_{k}\right)
g_{k-1}\left( \alpha_{g},n_{2},n_{3},...,n_{k-1}\right) ^{2}}{f_{k-1}\left(
\alpha_{g},n_{2},n_{3},...,n_{k-1}\right) }  \label{25}
\end{equation}
Substituting for $E_{j}$ from eq. (\ref{17}) into eqs. (\ref{22}) and (\ref%
{23}) we have%
\begin{equation}
f_{k-1}=1-\frac{\alpha_{g}^{2}}{2k}\sum_{j=2}^{k-1}\frac{\left( j-1\right)
^{3}}{jn_{j}^{2}}\frac{g_{j-1}^{3}}{f_{j-1}}  \label{26}
\end{equation}%
\begin{equation}
g_{k-1}=1-\frac{\alpha_{g}^{2}}{2\left( k-1\right) }\sum_{j=2}^{k-1}\frac{%
\left( j-1\right) ^{3}}{jn_{j}^{2}}\frac{g_{j-1}^{3}}{f_{j-1}}  \label{27}
\end{equation}

\subsection{Inequalities and asymptotics}

Since $r_{k}>0$ and $M_{k}>0$, for all $k$, we conclude from eqs. (\ref{18})
and (\ref{19}) that%
\begin{equation}
f_{k-1}>0\text{, }g_{k-1}>0  \label{28}
\end{equation}
From eqs. (\ref{26}) - (\ref{28}) we have%
\begin{equation}
f_{k-1}\leq1\text{, }g_{k-1}\leq1  \label{29}
\end{equation}
for all $k$ and $\alpha_{g}$. In this equation and subsequent ones the
equality holds if either $\alpha_{g}=0$ or $n_{2}=n_{3}=...=n_{k-1}=\infty$.
From (\ref{26}), and (\ref{27}) it trivially follows that%
\begin{equation}
f_{k-1}\geq g_{k-1}  \label{30}
\end{equation}
From eq. (\ref{29}) it follows that%
\begin{equation}
g_{k-1}\geq g_{k-1}^{2}\geq g_{k-1}^{3}  \label{31}
\end{equation}
From eqs. (\ref{30}) - (\ref{31}) we have%
\begin{equation}
f_{k-1}\geq g_{k-1}\geq g_{k-1}^{2}\geq g_{k-1}^{3}  \label{32}
\end{equation}
From the latter we conclude that%
\begin{equation}
\frac{g_{j-1}^{3}}{f_{j-1}}\leq1  \label{33}
\end{equation}
From eqs. (\ref{33}) and (\ref{17}) we have%
\begin{equation}
E_{k}\geq-\frac{\left( k-1\right) ^{3}}{2k}\frac{mc^{2}\alpha_{g}^{2}}{%
n_{k}^{2}}  \label{34}
\end{equation}
Letting \ \ $p=k\alpha_{g}$\ and \ $k\gg2$ \ in eq.(\ref{34}) we may rewrite
it as%
\begin{equation}
E_{k}\geq-\frac{p^{2}mc^{2}}{2n_{k}^{2}}  \label{35}
\end{equation}
If $p\rightarrow\infty$ as \ $k\rightarrow\infty$\ \ \ in eq.(\ref{35}),
then the energy of the system has no lower bound. Therefore the system will
go on losing energy during interactions until it becomes a part of the
infinite energy of the vacuum state of the universe. In this case, the
system cannot have an independent existence. This is averted if $p<M,$where\
M is a positive constant. In this case the system can have a stable
existence. Therefore \ it can be shown that $p<M$\ \ is the necessary and
sufficient condition for the stability of the Macro Holeum. In the following
we will assume $p$ to be a positive constant. Note that the right hand side
of eq.(\ref{35}) is identical in form to the equation for the energy
eigenvalues of a hydrogen atom with $m$ replacing the mass of the electron
and $p$ replacing the fine structure constant. Therefore it is clear that $p 
$ may be interpreted as an effective coupling constant for the Macro Holeum.

Now we consider the problem of reducing this many-body problem to a two-body
one. Here we are considering a truly gigantic system having a total mass
comparable to or even much greater than the solar mass. Here $k$ would be in
the astronomical range, say, $k=10^{50}$ to $k=10^{100}$. The huge mass of
this system gives it a huge inertia. It will be very difficult to perturb
such a system. Thus, we may assume that most of the system will remain
unperturbed and only the outermost one or two microscopic black holes will
be affected by the interactions of the system with its environment. In fact,
to reduce this many-body problem to a two-body one, we will make the extreme
simplification that all the microscopic black holes in the core are in the
same quantum state described by $n_{2}=n_{3}=...=n_{k-1}=n$, say, and that
the outermost microscopic black hole is in an arbitrary quantum state
described by the principal quantum number $n_{k}$. The first assumption
embodies the great inertia of the system.

We recall here that in the nuclear many-body problem, the nucleus is assumed
to be an infinite medium in which every nucleon moves in the same average
potential produced by the rest of them. This reduces the many-body problem
to a one-body problem. This, plus a heuristic spin-orbit potential with an
arbitrarily chosen sign, fetches us good quantitative agreement with nuclear
data in the nuclear shell model. Great complexity often yields to extreme
simplification; provided the simplification embodies the physics of the
system correctly. In our problem, if there is a greater excitation, we may
describe it in terms of $n,n_{k-1}$ and $n_{k}$; wherein the core now
consists of $k-2$ microscopic black holes all in the same state of
excitation $n$, and the two outer-most microscopic black holes are in
arbitrary states of excitation described by $n_{k-1}$ and $n_{k}$. But in
the first approximation, in the following, we describe the system in terms
of $n$ and $n_{k}$ only, apart from the other parameters.

Now we make a Taylor series expansion of $f_{k-1}$, eq. (\ref{26}), in
powers of $x=\alpha_{g}^{2}$. Then we get%
\begin{equation}
f_{k-1}(x)=1+xf_{k-1}^{\prime}(0)+\frac{x^{2}}{2!}f_{k-1}^{\prime\prime
}(0)+....  \label{36}
\end{equation}
From eq.(\ref{26}) we have%
\begin{equation}
f_{k-1}^{\prime}(x)=-\frac{1}{2k}\sum_{j=2}^{k-1}\frac{\left( j-1\right) ^{3}%
}{jn_{j}^{2}}\frac{g_{j-1}^{3}}{f_{j-1}}-\frac{x}{2k}\sum_{j=2}^{k-1}\frac{%
\left( j-1\right) ^{3}}{jn_{j}^{2}}\left[ \frac{3g_{j-1}^{2}g_{j-1}^{\prime}%
}{f_{j-1}}-\frac{g_{j-1}^{3}f_{j-1}^{\prime}}{f_{j-1}^{2}}\right]  \label{37}
\end{equation}
From eq.(\ref{37}) it follows that%
\begin{equation}
f_{k-1}^{\prime}(0)=-\frac{S_{k-1}}{2kn^{2}}  \label{38}
\end{equation}
In these equations we have taken $n_{2}=n_{3}=...=n_{k-1}=n.$ Similarly from
eq.(\ref{27}) it follows that%
\begin{equation}
g_{k-1}^{\prime}(0)=-\frac{S_{k-1}}{2(k-1)n^{2}}  \label{39}
\end{equation}
where%
\begin{align}
S_{k-1} & =\sum_{j=2}^{k-1}\frac{\left( j-1\right) ^{3}}{j}  \notag \\
& =\frac{\left( k-1\right) k\left( 2k-1\right) }{6}-\frac{3k\left(
k-1\right) }{2}+3\left( k-1\right) -C-\ln\left( k-1\right)  \notag \\
& -\frac{1}{2\left( k-1\right) }+\sum_{j=2}^{\infty}\frac{A_{j}}{\left(
k-1\right) k\left( k+1\right) ...\left( k+j-2\right) }  \label{40}
\end{align}
where $A_{2}=A_{3}=\frac{1}{12}$, $A_{4}=\frac{19}{80}$, etc. and $C$ is the
Euler constant; $C=0.577216$. Now substituting for $S_{k-1}$ from eq.(\ref%
{40}) into eq.(\ref{38}) and keeping only the two leading terms we get%
\begin{equation}
f_{k-1}^{\prime}(0)=-\frac{1}{2kn^{2}}\left[ \frac{k^{3}}{3}-\frac{3k^{2}}{2}%
+O(k)\right] ...  \label{41}
\end{equation}
\qquad\qquad\qquad\qquad\qquad\qquad\qquad\qquad\qquad\qquad%
\begin{align}
xf^{\prime}(0) & =-\frac{\alpha_{g}^{2}k^{2}}{6n^{2}}+\frac{3\alpha_{g}^{2}k%
}{4n^{2}}+O(\alpha_{g}^{2}k^{0})  \notag \\
& =-\frac{p^{2}}{6n^{2}}+\frac{3p^{2}}{4kn^{2}}+O(\frac{p^{2}}{k^{2}}) 
\notag \\
& =-\frac{p^{2}}{6n^{2}}+O(k^{-1})  \label{42}
\end{align}
\qquad\qquad\bigskip where $p=k\alpha_{g}$ is taken as a constant.
Differentiating eq.(\ref{37}) again with respect to $x$ and letting $x=0$ we
have%
\begin{equation}
f_{k-1}^{\prime\prime}(0)=-\frac{1}{kn^{2}}\sum_{j=2}^{k-1}\frac{\left(
j-1\right) ^{3}}{j}\left[ 3g_{k-1}^{\prime}(0)-f_{k-1}^{\prime}(0)\right]
\label{43}
\end{equation}
Substituting from eqs. (\ref{38}) and (\ref{39}) into eqs. (\ref{43}) we
have\bigskip%
\begin{equation}
f_{k-1}^{\prime\prime}(0)=-\frac{1}{2kn^{4}}\sum_{j=2}^{k-1}\frac {%
(2j+1)\left( j-1\right) ^{2}S_{j-1}}{j^{2}}  \label{44}
\end{equation}
\bigskip Since we need only the leading order terms for $k\gg2,$we
substitute only the first term of $\ S_{j-1}$ from eq. (\ref{40}) into eq. (%
\ref{44}) to get%
\begin{align}
f_{k-1}^{\prime\prime}(0) & =-\frac{1}{2kn^{4}}\sum_{j=2}^{k-1}\frac {%
(4j^{2}-1)\left( j-1\right) ^{3}}{6j}+l.o.t.  \notag \\
& =-\frac{1}{2kn^{4}}\sum_{j=2}^{k-1}\left[ \frac{2}{3}j^{4}-2j^{3}+\frac{11%
}{6}j^{2}-\frac{1}{6}j-\frac{1}{2}+\frac{1}{6j}\right] +l.o.t.  \label{45}
\end{align}
where $l.o.t.$ stands for lower order terms. Now we use the identity%
\begin{equation}
\sum_{i=1}^{m}i^{4}=\frac{m^{5}}{5}+\frac{m^{4}}{2}+\frac{m^{3}}{3}-\frac {m%
}{30}  \label{46}
\end{equation}
where $m=k-1\gg1.$Therefore substituting only the leading term from eq. (\ref%
{46}) into the leading term in eq. (\ref{45}) we get%
\begin{equation}
f_{k-1}^{\prime\prime}(0)=-\frac{k^{4}}{15n^{4}}+O(k^{3})  \label{47}
\end{equation}%
\begin{align}
\frac{x^{2}}{2!}f_{k-1}^{\prime\prime}(0) & =-\frac{k^{4}\alpha_{g}^{4}}{%
30n^{4}}+O(\frac{p^{4}}{k})  \notag \\
& =-\frac{p^{4}}{30n^{4}}+O(k^{-1})  \label{48}
\end{align}
Substituting eqs. (\ref{42}) and (\ref{48}) into eq. (\ref{36}) we get%
\begin{equation}
f_{k-1}(x)=1-\frac{p^{2}}{6n^{2}}-\frac{p^{4}}{30n^{4}}+O(k^{-1})...
\label{49}
\end{equation}
\bigskip Now we further assume $n\gg1.$This allows us to keep only the first
two terms in eq. (\ref{49}). Therefore we get%
\begin{equation}
f_{k-1}(x)\simeq1-\frac{p^{2}}{6n^{2}}+O(k^{-1})...  \label{50}
\end{equation}
Now for $k\gg2,$ $f_{k-1}$ $\simeq g_{k-1}$.Henceforth, for simplicity, we
will drop the notation $O\left( k^{-1}\right) $ and we will replace\ the
symbol $\simeq$ by $=$ \ in the following. Therefore, for $k\gg2$ and $n\gg1$
we have%
\begin{equation}
f_{k-1}=g_{k-1}=1-\frac{p^{2}}{6n^{2}}  \label{51}
\end{equation}
\bigskip Note that $1\geq f_{k-1}(x)\geq0$. Therefore from eq. (\ref{51}) we
have%
\begin{equation}
0\leq p^{2}\leq6.  \label{52}
\end{equation}
Substituting from eq. (\ref{51}) into eqs. (\ref{17}) - (\ref{20}) and eqs. (%
\ref{24}) and (\ref{25}), we have%
\begin{equation}
E_{k}=-\frac{p^{2}mc^{2}}{2n_{k}^{2}}\left( 1-\frac{p^{2}}{6n^{2}}\right)
^{2}  \label{53}
\end{equation}%
\begin{equation}
r_{k}=\frac{\pi^{2}kRn_{k}^{2}}{16p^{2}\left( 1-\frac{p^{2}}{6n^{2}}\right) }
\label{54}
\end{equation}%
\begin{equation}
M_{k}=mk\left( 1-\frac{p^{2}}{6n^{2}}\right)  \label{55}
\end{equation}%
\begin{equation}
R_{k}=kR\left( 1-\frac{p^{2}}{6n^{2}}\right)  \label{56}
\end{equation}%
\begin{equation}
\alpha_{k}=p\left( 1-\frac{p^{2}}{6n^{2}}\right)  \label{57}
\end{equation}%
\begin{equation}
\frac{R_{k}}{r_{k}}=\frac{16p^{2}}{\pi^{2}n_{k}^{2}}\left( 1-\frac{p^{2}}{%
6n^{2}}\right) ^{2}  \label{58}
\end{equation}
The density of the Macro Holeum is given by%
\begin{equation}
\rho_{k}=\frac{3072m_{p}p^{5}\left( 1-\frac{p^{2}}{6n^{2}}\right) ^{4}}{%
\pi^{7}kR_{p}^{3}n_{k}^{6}}  \label{59}
\end{equation}
In eqs. (\ref{53}) - (\ref{59}), $p=k\alpha_{g}$ satisfies eq. (\ref{52});
and $R_{p}$ given by%
\begin{equation}
R_{p}=\frac{2m_{p}G}{c^{2}}  \label{60}
\end{equation}
is the Schwarzschild radius of a microscopic black hole of Planck mass.
Apart from the integers $n$ and $n_{k}$ the seven properties listed in eqs. (%
\ref{53}) - (\ref{59}) depend, in general, on two parameters $m$ and $k$ or
their two suitable combinations. But $\alpha_{k}$ and $\frac{R_{k}}{r_{k}}$
depend only upon the combination $p=k\alpha_{g}$. As mentioned earlier, $p$
may be regarded as an effective coupling strength that determines the
formation of the Macro Holeum. The ratio $\frac{R_{k}}{r_{k}}$ determines
the class of the Macro Holeum as will be clear in the following.

\subsection{Classes of Holeum}

Eq. (\ref{52}) defines the physical region in which the bound state
formation can occur. This is because for $p^{2}>6$, $r_{k}$, $M_{k}$, $R_{k}$
and $\alpha_{k}$ become negative for some values of $n$. For $p^{2}\leq6$
this does not happen. Eq. (\ref{58}) enables us to classify the Macro
Holeums as follows:

\begin{enumerate}
\item If the ratio of the Schwarzschild radius $R_{k}$ of a Macro Holeum to
its bound state radius $r_{k}$, namely, $\frac{R_{k}}{r_{k}}$ is greater
than or equal to unity, then the Schwarzschild radius of the Macro Holeum is
greater than its physical radius . Therefore the Macro Holeum is a black
hole. It will emit Hawking radiation. We shall call such a Macro Holeum a 
\textbf{Black Holeum} and denote it as \textbf{BH}.

\item If the ratio in eq. (\ref{58}) is less than unity then $r_{k}>R_{k}$
and the bound state is not a black hole. It will not emit Hawking radiation
despite the fact that it consists of a large number of microscopic black
holes. We shall call it an \textbf{ordinary Macro Holeum} and denote it by 
\textbf{H}. We note that, in nature, a free neutron decays but a neutron in
a stable nucleus never does so. Therefore the fact that the microscopic
black holes inside the Macro Holeum H do not emit the Hawking radiation
should not come as a surprise to us.

\item From eq. (\ref{55}) we see that if $p=k\alpha_{g}=\sqrt{6}$ for $n=1$
then $M_{k}=0$ and the bound state will be a \emph{massless} Macro Holeum.
It will start moving at the speed of light as soon as it is produced,
subject to the conservation laws. We shall call it a \textbf{Lux Holeum} and
denote it by \textbf{LH}. Strictly speaking, we are not allowed to take $n=1$%
. Therefore, $p=\sqrt{6}$ may be taken as a nominal value and by keeping
more terms on the right hand side of eq. (\ref{49}), we may improve upon
this approximation to obtain a more accurate value of $p$. But it is always
true that we can put $M_{k}=0$ in eq. (\ref{55}), or more generally in eq.(%
\ref{19}). The LH is a bundle of gravitational energy which is presently
undetectable. Hence we may tentatively identify it with the Dark Energy
(DE). However, further theoretical development of the Holeum theory would
have to be awaited to see if the LHs possess the repulsive property
associated with the DE which is believed to be responsible for the expansion
of the universe.
\end{enumerate}

Now we would like to discuss this classification of Macro Holeums in more
quantitative detail. We rewrite eq. (\ref{58}) as follows:%
\begin{equation}
\frac{R_{k}}{r_{k}}=f\left( x\right) =ax\left( 1-bx\right) ^{2}  \label{61}
\end{equation}
where%
\begin{equation}
a=\frac{16}{\pi^{2}n_{k}^{2}}\text{, }b=\frac{1}{6n^{2}}\text{, }x=p^{2}
\label{62}
\end{equation}
Now $f\left( x\right) $ vanishes at $x=0,\frac{1}{b}$. It has a maximum at $%
x=\frac{1}{3b}=2n^{2}$ and a minimum at $x=\frac{1}{b}=6n^{2}$. If $f_{\max
}=f\left( \frac{1}{3b}\right) >1$, then the line $y=1$ intersects the curve $%
y=f\left( x\right) $ in three points $x_{1}$, $x_{2}$, $x_{3}$ such that%
\begin{align}
0 & <x_{1}<\frac{1}{3b}  \notag \\
\frac{1}{3b} & <x_{2}<\frac{1}{b}  \notag \\
\frac{1}{b} & <x_{3}<\infty  \label{63}
\end{align}
For illustrative purposes, taking $n_{k}=n=1$ we get $f_{\max}=1.441012>1$.
Note that in deriving the equations for the properties we have \ assumed $%
n\gg1$. Therefore, strictly speaking, we are not allowed to take $n=1$. But
for simplicity we do so. Similar results will be obtained for $n\gg1$ too.
The three values of $x_{i}$ such that $f\left( x_{i}\right) =1$ are found to
have the values $x_{1}=0.831192$, $x_{2}=3.470404$ and $x_{3}=7.698403$.
Thus, $f\left( x\right) <1$ for $0<x<x_{1}$ and we have H formation in this
range. For $x_{1}\leq x\leq x_{2}$ we have $f\left( x\right) \geq1$ and we
have BH formation. For $x_{2}<x<\frac{1}{b}$ we have $f\left( x\right) <1$
and we again have H formation. However since the latter parameter range is
higher than the earlier one and since there is a gap $[x_{1},x_{2}]$ between
them, we call the Macro Holeums in this range \textbf{Hyper Holeums} and
denote them as \textbf{HH}. Note that both H and HH are non-black hole
states with $\frac{R_{k}}{r_{k}}<1$. Hence these HH states, too; do not emit
Hawking radiation. The point \thinspace$x=6$ corresponds to the formation of
the Lux Holeums, LH for $n=1$. $x>6$ is an unphysical region because $r_{k}$%
, $M_{k}$, $R_{k}$ and $\alpha_{k}$ become negative in this range. Thus, for 
$f_{\max}>1$ we summarize the situation as follows:

\begin{equation}
\begin{tabular}{|l|l|}
\hline
$0<x<x_{1}$ & H \\ \hline
$x_{1}\leq x\leq x_{2}$ & BH \\ \hline
$x_{2}<x<\frac{1}{b}$ & HH \\ \hline
$x=\frac{1}{b}$ & LH \\ \hline
\end{tabular}
\label{64}
\end{equation}

On the other hand, for $f_{\max}<1$, the line $y=1$ intersects the curve $%
y=f\left( x\right) $ in only one point, namely, $x_{3}>\frac{1}{b}=6$ which
is in the unphysical region. In other words, we have $f\left( x\right) <1$
for the entire physical region $0<x<6$. In this case we have

\begin{equation}
\begin{tabular}{|l|l|}
\hline
$0<x<6$ & H \\ \hline
$x=6$ & LH \\ \hline
\end{tabular}
\label{65}
\end{equation}

There is no HH and BH formation in this case.

The formula for the gravitational energy spectrum of a Macro Holeum is
exactly the same for H, HH, and BH. It is given by eq. (\ref{53}). It is a
band spectrum. For a given value of $n$, $n_{k}$ takes all integral values
from $1$ to $\infty$. This gives rise to a band structure for each $n$.

\subsection{The Macro Holeum occupies space}

The most probable radius of a Macro Holeum is given by eq. (\ref{54}). From
the latter we see that a Macro Holeum is a layered structure. Every
microscopic black hole in it moves in its own separate orbit which is a
closed shell. No two orbits cross each other. From eqs. (\ref{52}) and (\ref%
{54}) it follows that%
\begin{equation}
r_{k}>\frac{\pi^{2}kRn_{k}^{2}}{16p^{2}}>\frac{\pi^{2}kRn_{k}^{2}}{96}
\label{66}
\end{equation}
This result implies that a Macro Holeum occupies space just like ordinary
matter. It cannot be compressed beyond the limit given on the right hand
side of eq. (\ref{66}). A similar result was derived for a di-Holeum where $%
k=2$ in \cite{Chavda.Chavda}. In the latter the formula for the radius was
derived by maximizing the probability density which is a purely quantum
mechanical concept without a classical analogue. Note that we are
considering spinless microscopic black holes. Therefore, this property of
non-compressibility is not to be confused with that following from the
anti-commutativity of fermionic operators, namely, the Pauli exclusion
principle. Note that naively and classically we could expect to get $%
r_{k}>kR $ from the non-overlap of the $k$ microscopic black holes in a
Macro Holeum. But here we get the far stronger result, eq. (\ref{66}), from
the quantum mechanical treatment. This means that the ground state of a
Macro Holeum can have a radius as small as $\frac{\pi^{2}}{96}$ or $10.3\%$
of its classical non-overlap radius. It cannot be compressed beyond that.
The presence of $n_{k}^{2}$ in the numerator and that of $1-\frac{p^{2}}{%
6n^{2}}$ in the denominator of eq. (\ref{54}) both greatly increase the
radius of a large Macro Holeum. This means that there is a lot of empty
space in a Macro Holeum. Therefore the density of Macro Holeum-type DM may
be considerably less than that of water, especially for very large Macro
Holeums; as we shall see shortly.

\subsection{The screening factor and the mass of a Macro Holeum}

A remarkable feature of the set of eqs. (\ref{53}) - (\ref{59}) is the
presence of, what one might call the "screening factor" $\left( 1-\frac {%
p^{2}}{6n^{2}}\right) $, in them. This factor depends upon $k$, $\alpha_{g}$
and $n$. The latter is the principal quantum number of each of the $k-1$
microscopic black holes constituting the "core" of the Macro Holeum.
Therefore it is the quantum mechanical orbital motion of the $k-1$ core
microscopic black holes which produces the antigravity-type reduction, $%
\frac{p^{2}}{6n^{2}}$ in the screening factor. There is another way to see
the significance of this term. We know that an isolated system eventually
settles down to its lowest energy state, i.e. its ground state. From eq. (%
\ref{53}) we see that the lowest energy state has $n=\infty$. This happens
only when $n_{2}=n_{3}=...n_{k-1}=\infty$. This is the case wherein all the
microscopic black holes constituting the core have been knocked out from
their orbits and are free to move randomly inside the orbit of the outermost
microscopic black hole. In this case the antigravity term $\frac{p^{2}}{%
6n_{c}^{2}}$, in the screening factor, vanishes. It is non-zero only when
there is orbital motion among the microscopic black holes of the core. This
screening factor has a dramatic effect on the mass $M_{k}$ of the Macro
Holeum, eq. (\ref{55}). If we take $p^{2}=6$ and $n=1$, then the screening
term vanishes and so does the mass of the Macro Holeum. This is the Lux
Holeum, LH. As discussed above, strictly speaking, we are not allowed to
take $n=1$, so this is a schematic argument. But we may take more terms in
eq. (\ref{49}) to find a more accurate value of $p$ such that $M_{k}=0$ to a
desired degree of accuracy.

\subsection{The Density}

The density of a Macro Holeum is given by eq. (\ref{59}). For $k\gg2$ it is
inversely proportional to $k$. It is also proportional to the fourth power
of the screening factor. Thus, it is clear that very massive Macro Holeums
would have very low density. From Tables $1$, $2$, and $3$ we see that for $%
k=10^{90}$ the densities of each of the Macro Holeums H and BH are of the
order of $10^{-4}$ times that of water whereas that of HH is a mere $10^{-8}$
times that of water and their masses are all greater than a million solar
masses. In other words, the supermassive Macro Holeums, including the
supermassive black holes existing today, have extremely low densities in
comparison with that of water. The density increases sharply as we go
towards the center of the Macro Holeum, but it always remains finite.

\subsection{The Energy Bands}

We expect the spectrum of energy eigenvalues of a Macro Holeum \ (H, HH, BH)
to be similar to that of a hydrogen atom. To see this in detail let us
recall that the energy eigenvalue of a hydrogen atom is given by%
\begin{equation}
E_{n}=-\frac{m_{e}c^{2}\alpha^{2}}{2n^{2}}  \label{67}
\end{equation}
where $m_{e}$ is the electron mass and $\alpha=\frac{e^{2}}{\hbar c}$ is the
fine structure constant representing the dimensionless coupling strength of
the electromagnetic interaction. We may rewrite eq. (\ref{53}) as:%
\begin{equation}
E_{k}=-\frac{mc^{2}\alpha_{k}^{2}}{2n_{k}^{2}}  \label{68}
\end{equation}
where $\alpha_{k}$ is given by eq. (\ref{57}). Thus, the energy eigenvalue
spectrum of a Macro Holeum is identical in form to that of a hydrogen atom.
However, there is a great surprise here. Naively, one would think that the
ground state, i.e. the state with the lowest energy, would be the one in
which $n_{k}=1$ and $n_{2}=n_{3}=...=n_{k-1}=1$. But a look at eq. (\ref{53}%
) negates this. From the latter we find that the state with the lowest
energy has $n_{k}=1$ and $n_{2}=n_{3}=...=n_{k-1}=\infty$. The latter fact
implies that all the $k-1$ microscopic black holes of the core have been
completely knocked out of their orbits and are free to move randomly inside
the orbit of the outermost microscopic black hole. Thus, the ground state of
the Macro Holeum is a bag full of randomly moving $k-1$ core microscopic
black holes being shepherded by a solitary "sentinel" microscopic black
hole, itself undergoing a quantized orbital motion characterized by a
principal quantum number $n_{k}$.

Although the energy spectra of a Macro Holeum and a hydrogen atom are very
similar, there are two important differences between them:

\begin{enumerate}
\item Whereas the hydrogen atom energy spectrum is a line spectrum, that of
the Macro Holeum is a band spectrum. This is because $n_{k}$ takes all
values from $1$ to $\infty$ for each value of $n$ in eq. (\ref{53}). This
gives rise to an energy band for each $n$.

\item Another important difference is that for a Macro Holeum, only those
transitions are allowed that obey $\Delta n_{k}=\pm2$. This is because the
graviton responsible for the gravitational interaction has a spin $J=2$ as
opposed to the spin $J=1$ of the photon responsible for the electromagnetic
interaction.
\end{enumerate}

The width of an energy band associated with a principal quantum number $%
n_{k} $ is given by%
\begin{equation}
\Delta\left( n_{k}\right) =\frac{p^{4}\left( 1-\frac{p^{2}}{12}\right) mc^{2}%
}{6n_{k}^{2}}  \label{69}
\end{equation}
This is obtained by keeping $n_{k}$ fixed and varying $n$ from $1$ to $%
\infty $. This equation shows that the broadest energy band belongs to the
state with $n_{k}=1$. In the latter case eq. (\ref{69}) gives the total
displacement suffered by the Lyman band.

\subsection{The determination of the parameters of a Macro Holeum from the
frequency band spectrum}

If the outermost microscopic black hole makes a transition from $n_{k}=n_{2}$
to $n_{k}=n_{1}$ while $n$ remains the same, then the Macro Holeum will emit
gravitational radiation of frequency $\nu$, say, given by%
\begin{equation*}
h\nu=h\nu_{R}\left( 1-\frac{p^{2}}{6n^{2}}\right) \left( \frac{1}{n_{1}^{2}}-%
\frac{1}{n_{2}^{2}}\right)
\end{equation*}%
\begin{equation}
h\nu_{R}=\frac{p^{2}mc^{2}}{2}  \label{70}
\end{equation}
Here $h\nu_{R}$ is the gravitational Rydberg constant.

Note that for a given value of $n$ we have the following spectrum:

\begin{enumerate}
\item Consider the transitions $n_{k}=1$ to $n_{k}=3,4,5...\infty$. This is
the gravitational analogue of the Lyman series of the hydrogen atom spectrum
except that the top line $L_{\alpha}$ corresponding to the transition $%
n_{k}=1$ to $n_{k}=2$ is absent due to the rule $\Delta\left( n_{k}\right)
=\pm2$ mentioned above.

\item Consider the transitions $n_{k}=2$ to $n_{k}=4,5,6...\infty$. This is
the gravitational analogue of the Balmer series of the hydrogen atom
spectrum except that the top line $B_{\alpha}$ corresponding to the
transition $n_{k}=2$ to $n_{k}=3$ is absent due to the reason already
mentioned.

\item The analogues of.the Paschen and the Brackett series follow the same
pattern.
\end{enumerate}

Each time, the top line is missing. This entire hydrogen atom-like spectrum
repeats itself for different values of $n=1,2,3...\infty$. Each time the
entire pattern shifts upward and there would be overlapping of bands
corresponding to different values of $n$. The $n=\infty$ spectrum rests at
the top of the pattern with the highest frequency, $\nu_{R},$ given by eq. (%
\ref{70}). Of course, the $n=\infty$ spectrum is the analogue of the line
spectrum of a hydrogen atom with the constraint $\Delta\left( n_{k}\right)
=\pm2$.

\subsection{The ball-park values for H, HH, BH}

We would like to study various properties such as the mass, the size, the
density and especially the frequency spectrum of the Macro Holeums H, HH,
and BH. In view of the fact that LIGO has started taking scientific data, it
is of utmost importance to find if these Macro Holeums emit gravitational
radiation in the LIGO frequency range, namely $1$ Hz to several kHz. And if
so, we would like to determine the ball-park values of the above-mentioned
properties of these Macro Holeums that emit the gravitational radiation in
the LIGO frequency range. To this end, we take $p=\frac{\pi}{4},\sqrt{2},$
and $\sqrt{5.8}$. These values correspond to Macro Holeums H, BH, and HH,
respectively. Note that all these properties can be calculated in terms of
just two parameters $p$ and $k$. For each value of $p$ we take a range of
values of $k$. The results are presented in Tables $1$, $2$, and $3$
respectively. Our main objective is to see if these Macro Holeums emit the
gravitational radiation in the LIGO frequency range. From these we see that
this is, indeed, the case: all three types of Macro Holeums radiate
gravitational radiation in the LIGO range if $k\geq10^{80}$. From Table 1,
we see that for $k=10^{80}$, the mass of the Macro Holeum H is about $87.02$
times the solar mass, the radius $r_{k}$ is about $319$ km, the density is
about $1.27\times10^{6}$ times that of water and the Rydberg frequency is
about $507$ Hz which is within the LIGO frequency range. Similarly for a
Macro Holeum BH, we see from Table 2 that the mass $M_{k}$ is about $86.77$
times the solar mass, the radius $r_{k}$ is about $178$ km, the density is
about $7.33\times10^{6}$ times that of water and the Rydberg frequency is
about $2.2$ kHz which is almost within the LIGO range. From Table 3 we see
that for a Macro Holeum HH, the mass $M_{k}$ is about $5.66$ solar masses,
the radius $r_{k}$ is about $1600$ km, the density is about $656$ times that
of water and the Rydberg frequency is $8.35$ kHz, which is outside the LIGO
range. But here if we take $k=10^{82}$ then the frequencies emitted by H,
BH, and HH lie between $50.71$ Hz and $834.87$ Hz. These frequencies are in
the LIGO range. The masses of the Macro Holeums emitting these frequencies
lie between $56.62$ solar masses and $870.24$ solar masses. Therefore the
ball-park value of $k$ is $10^{82}$. However, in view of the sweeping
assumption $n_{2}=n_{3}=...=n_{k-1}=n$, it would be prudent to take the
ball-park value in the range $k=10^{80}$ to $k=10^{85}$. The corresponding
properties of the Macro Holeums may be read off from Tables $1$, $2$, and $3$%
. By taking $k>10^{90}$ we can have frequencies in the milli-Hz range or
even smaller for all three classes of Macro Holeums.

We may determine these parameters by identifying the Lyman, the Balmer, the
Paschen, etc. bands and by measuring their frequencies. We need only two
inputs to determine all the properties. This also enables us to calculate
the ratio $\frac{R_{k}}{r_{k}}$ that determines the class H, BH, or HH.

\subsection{The "sentinel" mode and the line spectra}

It is necessary to devote special consideration to the "sentinel" mode or
the ground state of a Macro Holeum because it has unique properties.
Recalling that in this mode $n_{k}$ is arbitrary but $%
n_{2}=n_{3}=...=n_{k-1}=\infty$, we rewrite the equations for the properties
as follows:%
\begin{equation}
E_{k}=-\frac{p^{2}mc^{2}}{2n_{k}^{2}}=-\frac{h\nu_{R}}{n_{k}^{2}}  \label{71}
\end{equation}%
\begin{equation}
r_{k}=\frac{\pi^{2}kRn_{k}^{2}}{16p^{2}}  \label{72}
\end{equation}%
\begin{equation}
M_{k}=mk  \label{73}
\end{equation}%
\begin{equation}
R_{k}=kR  \label{74}
\end{equation}%
\begin{equation}
\alpha_{k}=p=k\alpha_{g}  \label{75}
\end{equation}%
\begin{equation}
\frac{R_{k}}{r_{k}}=\frac{16p^{2}}{\pi^{2}n_{k}^{2}}  \label{76}
\end{equation}
and%
\begin{equation}
\rho_{k}=\frac{3072m_{p}p^{5}}{\pi^{7}kR_{p}n_{k}^{6}}  \label{77}
\end{equation}
Although the eqs. (\ref{71}) - (\ref{77}) are derived from eqs. (\ref{53}) -
(\ref{59}) by letting $n=\infty$, they are exact in this limit. They can be
derived ab-initio without making any approximations. They can be used to
determine the Lyman, Balmer, etc. series for the earlier case which involves
considerable overlap. Eq. (\ref{71}) gives the energy eigenvalue of a
particle of mass $m$ orbiting a particle of infinite mass. Here $p$ plays
the role of an effective coupling constant. There is no restriction $%
p^{2}\leq6$ in this case because $r_{k}$, $M_{k}$, $R_{k}$ and $\alpha_{k}$
never become negative. There is no LH formation either. From eq. (\ref{71})
we see that now we have a pure line spectrum similar to that of the hydrogen
atom except that the restriction $\Delta n_{k}=\pm2$ still applies. Here,
too, the ratio $\frac{R_{k}}{r_{k}}$, given by eq. (\ref{76}), still
determines the class H or BH of the Macro Holeum. \ For example, if $\frac{%
R_{k}}{r_{k}}\geq1$, the Macro Holeum is a BH. Here we have $n_{k}\leq\frac{%
4p}{\pi}$. If $\frac{R_{k}}{r_{k}}<1$, we have an H state. Here we have $%
n_{k}>\frac{4p}{\pi}$. Thus as before, the same tower of excited states
contains BHs as the low-lying states and H states as the higher excited
states. In both H and BH cases, the frequency spectrum is given by%
\begin{equation}
h\nu=h\nu_{R}\left( \frac{1}{n_{1}^{2}}-\frac{1}{n_{2}^{2}}\right)
\label{78}
\end{equation}
which is the usual hydrogen atom line spectrum subject to $n_{2}=n_{1}\pm2$.
As in the previous case, it should be easy to determine the Rydberg constant 
$h\nu_{R}$ from the frequency spectrum. The identification and the
determination of the Lyman, Balmer, etc. series, would also enable us to
determine $p$ and $k$. All the other properties given by eqs. (\ref{71}) - (%
\ref{77}) as well as the frequency spectrum given by eq. (\ref{78}) can be
determined. This includes the determination of the class H or BH of the
Macro Holeums in this case.

\section{Implications of Holeum Theory}

We have listed a number of phenomena that can be attributed to Holeums in
the introduction and in \cite{Chavda.Chavda, Chavda.Chavda2} . In this
section we would like to add two more phenomena to this list. They are:

\begin{enumerate}
\item The internal structure of and the emission of gravitational radiation
by astrophysical black holes.

\item Cosmic Gravitational Background Radiation (CGBR).
\end{enumerate}

\subsection{Internal structure of black holes}

A black hole is characterized by two properties: the presence of an event
horizon and the emission of Hawking radiation. A Black Holeum, BH, described
above, has both of these properties; as has an astrophysical black hole
existing today. The latter may have a primordial origin or it may have been
produced as a result of the gravitational collapse of a star. But a black
hole is a black hole irrespective of its origin. For example, a hydrogen
atom may have been produced in a laboratory or it may be in a stellar
atmosphere or in an interstellar dust cloud. But all these hydrogen atoms
have the same physical and chemical properties and the same internal
structure. Thus, we will identify any black hole existing today with a Black
Holeum BH. We give in eqs. (\ref{53}) - (\ref{59}) all the properties of a
BH in terms of just two parameters $k$ and $p$; that can be determined from
the observed frequency spectrum of the gravitational radiation emitted by
it. Thus, the hypothesis that the existing black holes may be identified
with BHs enables us to predict that all existing black holes have internal
structure and that they emit quantized gravitational radiation. This can be
tested. Of course, if the black hole is a part of a binary system, then it
will also emit classical gravitational radiation predicted by the general
theory of relativity. This is continuous radiation whereas the Holeum theory
predicts either a band spectrum or a line spectrum. The Holeum theory puts
the black holes in the larger class of Dark Matter objects, namely, Macro
Holeums BH.

\subsection{Cosmic Gravitational Background Radiation}

Although the universe is inhomogeneous locally, it is homogeneous on a large
scale of the order of $100$ Mpc. This means that apart from the
gravitational radiation from the nearby Macro Holeums contained in a sphere
of radius $100$ Mpc, the gravitational radiation received by LIGO would be
uniform in all directions. If we could filter out the radiation from all
Macro Holeums in a sphere of radius $100$ Mpc we will be left with a uniform
gravitational radiation from all directions. We call it Cosmic Gravitational
Background Radiation (CGBR). Similar Cosmic Microwave Background Radiation
(CMBR) was observed by Robert Wilson and Arno Penzias in 1964 \cite%
{PenziasWilson}. The CGBR arises from the homogeneity of space on a large
scale. The CMBR, \ on the other hand, is a remnant of the Big Bang.

\section{Discussion and Conclusions}

We have presented a model of a gravitational bound state of a very large
number of microscopic black holes of primordial or non-primordial origin. We
call it a Macro Holeum. It emits quantized gravitational radiation. The huge
inertia inherent in the system is used to reduce this many-body problem to a
two-body one. There are four classes of a Macro Holeum: H, BH, HH and LH
depending upon the mass of the constituents. There are five noteworthy
properties of a Macro Holeum:

\begin{enumerate}
\item We know that the ordinary fermionic matter occupies space. Here we
have shown that the bosonic Macro Holeums also occupy space. A Macro Holeum
can not be compressed beyond a certain limit. This is a quantum mechanical
result not related to the Pauli Exclusion Principle that applies only to
fermions. In other words, the Holeum and the Macro Holeum type bosonic DM
shares one very important property with the ordinary fermionic matter,
namely, it too occupies space.

\item It is well-known that natural radioactivity can not be switched off.
Yet the microscopic black holes in H and HH do not emit the Hawking
radiation that they would emit in their free state. This may look
surprising, but it is not; for the neutrons in stable nuclei do not emit
their beta radioactivity.

\item The orbital motions of the large number of microscopic black holes
produce an anti-gravity type effect that affects all the properties of a
Macro Holeum. In particular, we can vary the mass of a Macro Holeum by
manipulating the orbital excitations of its constituents. And, in fact, by
taking $p=\sqrt{6}$ with $n=1,$we can create a massless LH.

\item On general grounds, we prove that a Macro Holeum is a stable quantum
system just like a hydrogen atom.

\item All massive Macro Holeums H, BH, HH emit a hydrogen-like spectrum of
gravitational radiation with some overlapping of bands. The two parameters $%
m $ and $k$ can be determined from the spectrum. A gratifying feature of
this theory is the set of equations (\ref{71}) - (\ref{78}) for the
properties of the ground state. These are exact and can be derived directly
without making the simplifications and the approximations of the theory.
They can be used as a guide to figure out the spectral details and thus to
determine $m$ and $k$. A surprising prediction of the theory is that even
stationary, non-rotating, stand-alone astrophysical black holes emit
quantized gravitational radiation. We expect that such astrophysical black
holes and other Macro Holeums having their masses in the ball-park range of $%
57$ to $870$ solar masses would be emitting quantized gravitational
radiation in the LIGO frequency range.
\end{enumerate}

From Tables $1$ and $2$ we see that H and BH have very similar properties.
Since we have identified astrophysical black holes with the BHs, there is a
real possibility that an ordinary Macro Holeum H may be misidentified as an
astrophysical black hole existing today. This is because both Hs and BHs are
invisible and both would form accretion discs around themselves when they
are in binaries with visible stars. An ordinary Holeum H would permit light
to pass through it ; albeit after refracting it whereas a black hole has an
event horizon and is completely opaque. One way to distinguish between the
two would be to analyze their gravitational spectra that, as we have seen,
allow us to completely characterize the source. But at the present time
there is no detector that can target a specified source. Meanwhile it is
necessary to bear in mind that some of the objects identified as black holes
may, in fact, be ordinary Macro Holeums H.

If Macro Holeums and black holes having their masses in the ball-park range
of $57$ to $870$ solar masses are common-place in the universe, then the
prospects of LIGO detecting their gravitational radiation would be bright.
On this reckoning one would not be surprised if LIGO detects the
gravitational radiation from Macro Holeums and black holes more readily than
from the sources it was built to detect, namely, the classical gravitational
radiation predicted by the General Theory of Relativity.

This is a theory of DM. It is equally well a theory of gravitational
radiation. As mentioned in \cite{Chavda.Chavda}, it has been developed in
response to the call issued at the Rome conference on gravitational waves to
the theoretical physicists to make new predictions that can be tested at the
upcoming facilities like LIGO, VIRGO, etc. Now that LIGO is operational, we
hope it will give its verdict soon.

\section{Tables}

\subsection{Properties of a Macro Holeum, H}

\begin{tabular}{|l|l|l|l|l|l|}
\hline
$\mathbf{k}$ & $\mathbf{m}$\textbf{\ GeV} & $\mathbf{h\nu}_{R}$\textbf{\ Hz}
& $\mathbf{r}_{k}$\textbf{\ cm} & $\mathbf{M}_{k}\mathbf{/M}_{\odot}$ & $%
\mathbf{\rho}_{k}$\textbf{\ g/cm}$^{3}$ \\ \hline
$10^{50}$ & $1.0822\times10^{-6}$ & $5.0706\times10^{17}$ & $3.1925\times
10^{-8}$ & $8.7024\times10^{-14}$ & $1.2699\times10^{36}$ \\ \hline
$10^{60}$ & $1.0822\times10^{-11}$ & $5.0706\times10^{12}$ & $3.1925\times
10^{-3}$ & $8.7024\times10^{-9}$ & $1.2699\times10^{26}$ \\ \hline
$10^{70}$ & $1.0822\times10^{-16}$ & $5.0706\times10^{7}$ & $3.1925\times
10^{2}$ & $8.7024\times10^{-4}$ & $1.2699\times10^{16}$ \\ \hline
$10^{72}$ & $1.0822\times10^{-17}$ & $5.0706\times10^{6}$ & $3.1925\times
10^{3}$ & $8.7024\times10^{-3}$ & $1.2699\times10^{14}$ \\ \hline
$10^{74}$ & $1.0822\times10^{-18}$ & $5.0706\times10^{5}$ & $3.1925\times
10^{4}$ & $8.7024\times10^{-2}$ & $1.2699\times10^{12}$ \\ \hline
$10^{76}$ & $1.0822\times10^{-19}$ & $5.0706\times10^{4}$ & $3.1925\times
10^{5}$ & $8.7024\times10^{-1}$ & $1.2699\times10^{10}$ \\ \hline
$10^{78}$ & $1.0822\times10^{-20}$ & $5.0706\times10^{3}$ & $3.1925\times
10^{6}$ & $8.7024\times10^{0}$ & $1.2699\times10^{8}$ \\ \hline
$10^{80}$ & $1.0822\times10^{-21}$ & $5.0706\times10^{2}$ & $3.1925\times
10^{7}$ & $8.7024\times10^{1}$ & $1.2699\times10^{6}$ \\ \hline
$10^{82}$ & $1.0822\times10^{-22}$ & $5.0706\times10^{1}$ & $3.1925\times
10^{8}$ & $8.7024\times10^{2}$ & $1.2699\times10^{4}$ \\ \hline
$10^{84}$ & $1.0822\times10^{-23}$ & $5.0706\times10^{0}$ & $3.1925\times
10^{9}$ & $8.7024\times10^{3}$ & $1.2699\times10^{2}$ \\ \hline
$10^{86}$ & $1.0822\times10^{-24}$ & $5.0706\times10^{-1}$ & $3.1925\times
10^{10}$ & $8.7024\times10^{4}$ & $1.2699\times10^{0}$ \\ \hline
$10^{88}$ & $1.0822\times10^{-25}$ & $5.0706\times10^{-2}$ & $3.1925\times
10^{11}$ & $8.7024\times10^{5}$ & $1.2699\times10^{-2}$ \\ \hline
$10^{90}$ & $1.0822\times10^{-26}$ & $5.0706\times10^{-3}$ & $3.1925\times
10^{12}$ & $8.7024\times10^{6}$ & $1.2699\times10^{-4}$ \\ \hline
\end{tabular}

\smallskip Here $p=\frac{\pi}{4}$, $n=n_{k}=1$, except for $\rho_{k}$ where $%
n_{k}=10$ and $n=1$. $\frac{R_{k}}{r_{k}}=0.80501$. $M_{\odot}$ is the mass
of the sun.

\subsection{Properties of a Macro Holeum BH}

\begin{tabular}{|l|l|l|l|l|l|}
\hline
$\mathbf{k}$ & $\mathbf{m}$\textbf{\ GeV} & $\mathbf{h\nu}_{R}$\textbf{\ Hz}
& $\mathbf{r}_{k}$\textbf{\ cm} & $\mathbf{M}_{k}\mathbf{/M}_{\odot}$ & $%
\mathbf{\rho}_{k}$\textbf{\ g/cm}$^{3}$ \\ \hline
$10^{50}$ & $1.4521x10^{-6}$ & $2.2061x10^{18}$ & $1.7782x10^{-8}$ & $%
8.6767x10^{-14}$ & $7.3282x10^{36}$ \\ \hline
$10^{60}$ & $1.4521x10^{-11}$ & $2.2061x10^{13}$ & $1.7782x10^{-3}$ & $%
8.6767x10^{-9}$ & $7.3282x10^{26}$ \\ \hline
$10^{70}$ & $1.4521x10^{-16}$ & $2.2061x10^{8}$ & $1.7782x10^{2}$ & $%
8.6767x10^{-4}$ & $7.3282x10^{16}$ \\ \hline
$10^{72}$ & $1.4521x10^{-17}$ & $2.2061x10^{7}$ & $1.7782x10^{3}$ & $%
8.6767x10^{-3}$ & $7.3282x10^{14}$ \\ \hline
$10^{74}$ & $1.4521x10^{-18}$ & $2.2061x10^{6}$ & $1.7782x10^{4}$ & $%
8.6767x10^{-2}$ & $7.3282x10^{12}$ \\ \hline
$10^{76}$ & $1.4521x10^{-19}$ & $2.2061x10^{5}$ & $1.7782x10^{5}$ & $%
8.6767x10^{-1}$ & $7.3282x10^{10}$ \\ \hline
$10^{78}$ & $1.4521x10^{-20}$ & $2.2061x10^{4}$ & $1.7782x10^{6}$ & $%
8.6767x10^{0}$ & $7.3282x10^{8}$ \\ \hline
$10^{80}$ & $1.4521x10^{-21}$ & $2.2061x10^{3}$ & $1.7782x10^{7}$ & $%
8.6767x10^{1}$ & $7.3282x10^{6}$ \\ \hline
$10^{82}$ & $1.4521x10^{-22}$ & $2.2061x10^{2}$ & $1.7782x10^{8}$ & $%
8.6767x10^{2}$ & $7.3282x10^{4}$ \\ \hline
$10^{84}$ & $1.4521x10^{-23}$ & $2.2061x10^{1}$ & $1.7782x10^{9}$ & $%
8.6767x10^{3}$ & $7.3282x10^{2}$ \\ \hline
$10^{86}$ & $1.4521x10^{-24}$ & $2.2061x10^{0}$ & $1.7782x10^{10}$ & $%
8.6767x10^{4}$ & $7.3282x10^{0}$ \\ \hline
$10^{88}$ & $1.4521x10^{-25}$ & $2.2061x10^{-1}$ & $1.7782x10^{11}$ & $%
8.6767x10^{5}$ & $7.3282x10^{-2}$ \\ \hline
$10^{90}$ & $1.4521x10^{-26}$ & $2.2061x10^{-2}$ & $1.7782x10^{12}$ & $%
8.6767x10^{6}$ & $7.3282x10^{-4}$ \\ \hline
\end{tabular}

\smallskip Here $p^{2}=2,$ $n=n_{k}=1$, except for $\rho_{k}$ where $%
n_{k}=10 $ and $n=1$. $\frac{R_{k}}{r_{k}}=1.4410.$ $M_{\odot}$ is the mass
of the sun.

\subsection{Properties of a Macro Holeum HH}

\begin{tabular}{|l|l|l|l|l|l|}
\hline
$\mathbf{k}$ & $\mathbf{m}$\textbf{\ GeV} & $\mathbf{h\nu}_{R}$\textbf{\ Hz}
& $\mathbf{r}_{k}$\textbf{\ cm} & $\mathbf{M}_{k}\mathbf{/M}_{\odot}$ & $%
\mathbf{\rho}_{k}$\textbf{\ g/cm}$^{3}$ \\ \hline
$10^{50}$ & $1.8950x10^{-6}$ & $8.3487x10^{18}$ & $1.6003x10^{-7}$ & $%
5.6616x10^{-19}$ & $6.5595x10^{32}$ \\ \hline
$10^{60}$ & $1.8950x10^{-11}$ & $8.3487x10^{13}$ & $1.6003x10^{-2}$ & $%
5.6616x10^{-10}$ & $6.5595x10^{22}$ \\ \hline
$10^{70}$ & $1.8950x10^{-16}$ & $8.3487x10^{8}$ & $1.6003x10^{3}$ & $%
5.6616x10^{-5}$ & $6.5595x10^{12}$ \\ \hline
$10^{72}$ & $1.8950x10^{-17}$ & $8.3487x10^{7}$ & $1.6003x10^{4}$ & $%
5.6616x10^{-4}$ & $6.5595x10^{10}$ \\ \hline
$10^{74}$ & $1.8950x10^{-18}$ & $8.3487x10^{6}$ & $1.6003x10^{5}$ & $%
5.6616x10^{-3}$ & $6.5595x10^{8}$ \\ \hline
$10^{76}$ & $1.8950x10^{-19}$ & $8.3487x10^{5}$ & $1.6003x10^{6}$ & $%
5.6616x10^{-2}$ & $6.5595x10^{6}$ \\ \hline
$10^{78}$ & $1.8950x10^{-20}$ & $8.3487x10^{4}$ & $1.6003x10^{7}$ & $%
5.6616x10^{-1}$ & $6.5595x10^{4}$ \\ \hline
$10^{80}$ & $1.8950x10^{-21}$ & $8.3487x10^{3}$ & $1.6003x10^{8}$ & $%
5.6616x10^{0}$ & $6.5595x10^{2}$ \\ \hline
$10^{82}$ & $1.8950x10^{-22}$ & $8.3487x10^{2}$ & $1.6003x10^{9}$ & $%
5.6616x10^{1}$ & $6.5595x10^{0}$ \\ \hline
$10^{84}$ & $1.8950x10^{-23}$ & $8.3487x10^{1}$ & $1.6003x10^{10}$ & $%
5.6616x10^{2}$ & $6.5595x10^{-2}$ \\ \hline
$10^{86}$ & $1.8950x10^{-24}$ & $8.3487x10^{0}$ & $1.6003x10^{11}$ & $%
5.6616x10^{3}$ & $6.5595x10^{-4}$ \\ \hline
$10^{88}$ & $1.8950x10^{-25}$ & $8.3487x10^{-1}$ & $1.6003x10^{12}$ & $%
5.6616x10^{4}$ & $6.5595x10^{-6}$ \\ \hline
$10^{90}$ & $1.8950x10^{-26}$ & $8.3487x10^{-2}$ & $1.6003x10^{13}$ & $%
5.6616x10^{5}$ & $6.5595x10^{-8}$ \\ \hline
\end{tabular}

\smallskip Here $p^{2}=5.8,$ $n=n_{k}=1$, except for $\rho _{k}$ where $%
n_{k}=10$ and $n=1$. $\frac{R_{k}}{r_{k}}=1.0447\times 10^{-2}.$ $M_{\odot }$
is the mass of the sun.

\end{document}